\documentclass[12pt,english]{article}
\usepackage{bookman}
\usepackage[T1]{fontenc}
\usepackage[latin1]{inputenc}
\usepackage{amsmath}
\usepackage{graphicx}
\usepackage{amssymb}

\makeatletter

\providecommand{\LyX}{L\kern-.1667em\lower.25em\hbox{Y}\kern-.125emX\@}

\def\fnum@table{\tablename~{\bf\thetable}}
\def\fnum@figure{\figurename~{\bf\thefigure}}
\def\tablename{\footnotesize{\bf Table}}
\def\figurename{\footnotesize{\bf Figure}}

\usepackage{babel}
\makeatother
\begin{document}

\title{\textbf{Comparison of Microcanonical and Canonical Hadronization}}

\author{F.M. Liu$^{1,2,3,\, }$%
\footnote{Alexander von Humboldt Fellow%
}, K. Werner$^{3}$, J. Aichelin$^{3}$}

\maketitle
\noindent $^{1}$ \emph{\small Institute of Particle Physics, Central
China Normal University, Wuhan, China. }{\small \par}

\noindent \emph{\small $^{2}$ Institut fuer Theoretische Physik,
JWG Frankfurt Universitaet, Germany }{\small \par}

\noindent \emph{\small $^{3}$ Laboratoire SUBATECH, University of
Nantes - IN2P3/CNRS - Ecole des Mines de Nantes, Nantes, France}{\small \par}

\begin{abstract}
Average multiplicities and transverse momenta of hadrons are calculated
using a microcanonical hadronization description for a cluster of
given total energy and volume. As a function of the total energy,
we determine the critical volume above which the microcanonical description
coincides with the canonical one, and compare the results with those
obtained using one of standard canonical models. We show that the
critical volume depends on the energy and the mass of the hadrons.
For heavy particles, volumes above $50\, fm^{3}$ are needed, even
more than$100\, fm^{3}$ if one considers transverse momenta. Thus
the prediction of heavy hadron multiplicities in pp, Kp, and $e^{+}e^{-}$
reactions requires a microcanonical approach, whereas for heavy ion
reactions a canonical calculation is valid. We conclude by showing
the importance of the feeding for the observed hadron multiplicities. 
\end{abstract}

\section{Introduction}

The statistical description of proton-proton and heavy-ion reactions
has already a long tradition. It was Hagedorn \cite{hag1,hag2}, who
showed in a sequence of papers that many aspects of these reactions
are close to that one expects assuming that the transition matrix
element is constant and therefore the distribution of final state
particles is completely determined by phase space. 

More recently the statistical interpretation of nuclear reactions
regained interest, after it had been demonstrated that in heavy ion
reactions at CERN--SPS amd RHIC energies the multiplicities of a multitude
of non strange hadrons is in remarkable agreement with the assumption
that all particles are created in thermal equilibrium at a temperature
which is close to that expected from lattice calculations for the
transition of a hadron gaz towards a quark-gluon plasma {[}3-5{]}.
Also strange hadrons fit into this picture if one includes a penalty
factor for each strange quark which is contained in a hadron. 

Later this approach has been successfully extended towards AGS and
SIS energies \cite{ker}, as well as towards pp and $e^{+}e^{-}$
reactions \cite{becpp,becee}. Whereas the former are large systems
with a large reaction volume, the latter ones yield only a small hadron
multiplicity and require volumes of the order of 25 $fm^{3}$. For
such small systems, it is not evident that a canonical or a grand
canonical description is justified.

Therefore it is worthwhile to check whether for these small volumes
and particle multiplicities a microcanonical description still coincides
with the canonical one. It is the purpose of this article to investigate
this question employing for the first time a numerical realization
of the algorithm to calculate the microcanonical phase space which
has been presented by us in ref. \cite{wer}. For the present investigation,
we limit the number of hadrons to 54, which include pseudoscalar and
vector mesons (octet and singlet) as well as the octet and decouplet
of baryons and antibaryons. Strange particles are produced according
to phase space, i.e. without applying any suppression factor. The
results are compared to that of two canonical calculations using the
approach of Becattini et al. \cite{bec1,becpp,becee}, but without
strangeness suppression: in the first calculation the number of hadrons
which can be produced is limited to the same 54 species allowed in
our microcanonical treatment, in the second calculation the standard
set of hadrons \cite{bec1,becpp,becee} is included. In principle
we can include more hadrons in the microcanonical ensemble. This makes
a detailed comparison more difficult, because the less known decay
channels of the additional hadrons have to agree.

\section{Microcanonical Calculation}

Following the general philosophy of statistical approaches to hadron
production, we suppose that the result of a high energy collision
can be considered as a distribution of {}``clusters'', {}``droplets'',
or {}``fireballs'', which move relative to each other. Here, we
are only interested in $4\pi $ particle yields and average transverse
momenta, and therefore collective longitudinal motion needs not to
be considered, and the distribution of clusters may be identified
with one single {}``equivalent cluster'', being characterized by
its volume $V$ (the sum of individual proper volumes), its energy
$E$ (the sum of all the cluster masses) , and the net flavor content
$Q=(N_{u}-N_{\bar{u}},N_{d}-N_{\bar{d}},N_{s}-N_{\bar{s}})$. 

The basic assumption is that a cluster, characterized by $V$, $E$,
and $Q$, decays {}``statistically'' according to phase space. More
precisely, the probability of a cluster to hadronize into a configuration
$K=\{h_{1},\ldots ,h_{n}\}$ of hadrons $h_{i}$ is given by the microcanonical
partition function $\Omega (K)$ of an ideal, relativistic gas of
the $n$ hadrons $h_{i}$ \cite{wer},\[
\Omega (K)=\frac{V^{n}}{(2\pi \hbar )^{3n}}\, \prod _{i=1}^{n}g_{i}\, \prod _{\alpha \in \mathcal{S}}\, \frac{1}{n_{\alpha }!}\, \int \prod _{i=1}^{n}d^{3}p_{i}\, \delta (E-\Sigma \varepsilon _{i})\, \delta (\Sigma \vec{p}_{i})\, \delta _{Q,\Sigma q_{i}},\]
with $\varepsilon _{i}=\sqrt{m_{i}^{2}+p_{i}^{2}}$ being the energy,
and $\vec{p}_{i}$ the 3-momentum of particle $i$. The term $\delta _{Q,\Sigma q_{i}}$
ensures flavour conservation; $q_{i}$ is the flavour vector of hadron
$i$. The symbol $\mathcal{S}$ represents the set of hadron species
considered: we take $\mathcal{S}$ to contain the pseudoscalar and
vector mesons $(\pi ,K,\eta ,\eta ',\rho ,K^{*},\omega ,\phi )$ and
the lowest spin-$\frac{1}{2}$ and spin-$\frac{3}{2}$ baryons $(N,\Lambda ,\Sigma ,\Xi ,\Delta ,\Sigma ^{*},\Xi ^{*},\Omega )$
and the corresponding antibaryons. $n_{\alpha }$ is the number of
hadrons of species $\alpha $, and $g_{i}$ is the degeneracy of particle
$i$.

We are going to employ Monte Carlo techniques, so we have to generate
randomly configurations $K$ according to the probability distribution
$\Omega (K)$. We need a method in particular for intermediate size
droplets, covering droplet masses from few GeV up to 100 or 1000 GeV.
So the method should work for particle numbers $n=|K|$ between 2
and $10^{3}$, which means, we have to deal with a huge configuration
space. Such problems are well known in statistical physics, and the
method at hand is to construct a Markov process. So for a given cluster
with mass $E$, volume $V$, and flavor $Q$, we start from some arbitrary
initial configuration $K_{0}$, and generate a sequence $K_{0},K_{1},\ldots ,K_{I_{\textrm{eq}}}$,
with $I_{\textrm{eq}}$ being sufficiently large to have reached equilibrium
(which is defined to be the steady state of the Markov process). If
we repeat this procedure many times, getting configurations $K_{I_{\textrm{eq}}}^{(1)},K_{I_{\textrm{eq}}}^{(2)},\ldots ,$
these configurations are distributed as $\Omega (K)$. We need a transition
probability $p$ such that it leads to an equilibrium distribution
$\Omega (K)$, with the initial transient $I_{\textrm{eq}}$ being
as small as possible. Such an algorithm has been realized for the
first time in \cite{wer}.

The problem is solved in several steps. One first writes the phase
space integral as \begin{eqnarray}
 &  & \phi (E,m_{1},\ldots ,m_{n})=\int \prod _{i=1}^{n}d^{3}p_{i}\, \delta (E-\Sigma \varepsilon _{i})\, \delta (\Sigma \vec{p}_{i})\label{7}\\
 &  & =(4\pi )^{n}\int \prod _{i=1}^{n}dp_{i}\prod _{i=1}^{n}p_{i}^{2}\, \delta (E-\sum _{i=1}^{n}\varepsilon _{i})\, W(p_{1},\ldots ,p_{n}),\nonumber 
\end{eqnarray}
 with $p_{i}=|\vec{p}_{i}|$, and with the {}``random walk function''
$W$ given as \begin{equation}
W(p_{1},\ldots ,p_{n}):=\frac{1}{(4\pi )^{n}}\int \prod _{i=1}^{n}d\Omega _{i}\, \delta \big (\sum _{i=1}^{n}p_{i}\hat{e}_{i}\big ),\label{8}\end{equation}
 with $\hat{e}_{i}=\vec{p}_{i}/|\vec{p}_{i}|$. The name {}``random
walk function'' is due to the fact that $W$ represents the probability
to return back to the origin after $n$ {}``random walks'' $p_{i}\hat{e}_{i}$
with given step sizes $p_{i}$. In \cite{wer}, many details can be
found about an efficient calculation of $W$ for any $n$ (big or
small).

The next step amounts to getting rid of the energy delta function.
A variable transformation gives

\begin{eqnarray}
 &  & \phi (E,m_{1},\ldots ,m_{n})\label{47}\\
 &  & =\int _{0}^{1}dr_{1}\ldots \int _{0}^{1}dr_{n-1}\, \psi (E,m_{1},\ldots ,m_{n};r_{1},\ldots ,r_{n-1}),\nonumber 
\end{eqnarray}
 with \begin{eqnarray}
 &  & \psi (E,m_{1},\ldots ,m_{n};r_{1},\ldots ,r_{n-1})\label{48}\\
 &  & =\frac{(4\pi )^{n}\, T^{n-1}}{(n-1)!}\prod _{i=1}^{n}p_{i}\, \varepsilon _{i}\, W(p_{1},\ldots ,p_{n}).\nonumber 
\end{eqnarray}
 The symbol $T$ denotes the total kinetic energy $E-\Sigma m_{i}$,
and the absolute values of the momenta are expressed in terms of the
$r_{i}$ as \begin{eqnarray}
p_{i} & = & \sqrt{t_{i}(t_{i}+2m_{i})}\label{49}\nonumber \\
t_{i} & = & T(x_{i}-x_{i-1}),\, \, x_{0}=0\nonumber \\
x_{i} & = & x_{i+1}\sqrt[i]{r_{i}},\, \, x_{n}=1.\label{49}
\end{eqnarray}

In principle one may use Monte Carlo techniques to calculate the integral,
but this is very time consuming. A more elegant method amounts to
generalizing the hadronic final state by considering not only hadron
species, but also their momenta. So we use the generalized configurations
\begin{equation}
G=\{h_{1},\ldots ,h_{n};r_{1},\ldots ,r_{n-1}\},\end{equation}
where the $r_{i}$ are related to the momenta $p_{i}$ via eq. (\ref{49}).
The weight of such a configuration is \[
\Omega (G)=\frac{V^{n}}{(2\pi \hbar )^{3n}}\, \prod _{i=1}^{n}g_{i}\, \prod _{\alpha \in \mathcal{S}}\, \frac{1}{n_{\alpha }!}\, \, \psi (E,m_{1},\ldots ,m_{n};r_{1},\ldots ,r_{n-1}).\]
This expression is well suited to generate configurations $G$ according
to $\Omega (G)$, by constructing Markov chains, for details see \cite{wer}.

Flavor conservation is trivial to take into account, by considering
only propositions in the Markov chain construction which conserve
the total flavor $Q.$

Our algorithm provides a fast method to generate hadron configurations,
characterized by the number of hadrons, their type, and their momenta.
In this sense we have a real {}``event generator'' of statistically
generated hadrons.

\section{Results}

\subsection{Multiplicities}

In order to see best how the results of a microcanonical calculation
approach those of a canonical one, we investigate the average particle
density $\rho _{\alpha }=n_{\alpha }/V$, where $n_{\alpha }$ is
the average particle multiplicity $n_{\alpha }$ of hadron species
$\alpha $, and $V$ is the volume. In the microcanoncial approach,
the average particle density is in general a function of the two variables
$E$ and $V$. For large volumes, however, $\rho $ should only depend
on the ratio $\varepsilon =E/V$, and this represents the limit where
microcanonical and canonical description should coincide. 

We therefore calculate the density $\rho _{\alpha }$ as a function
of the energy density $\varepsilon $, for different volumes. We expect
these curves to converge for large volumes.%
\begin{figure}[htb]
\begin{flushleft}\vspace{-2cm}\includegraphics[  scale=0.7]{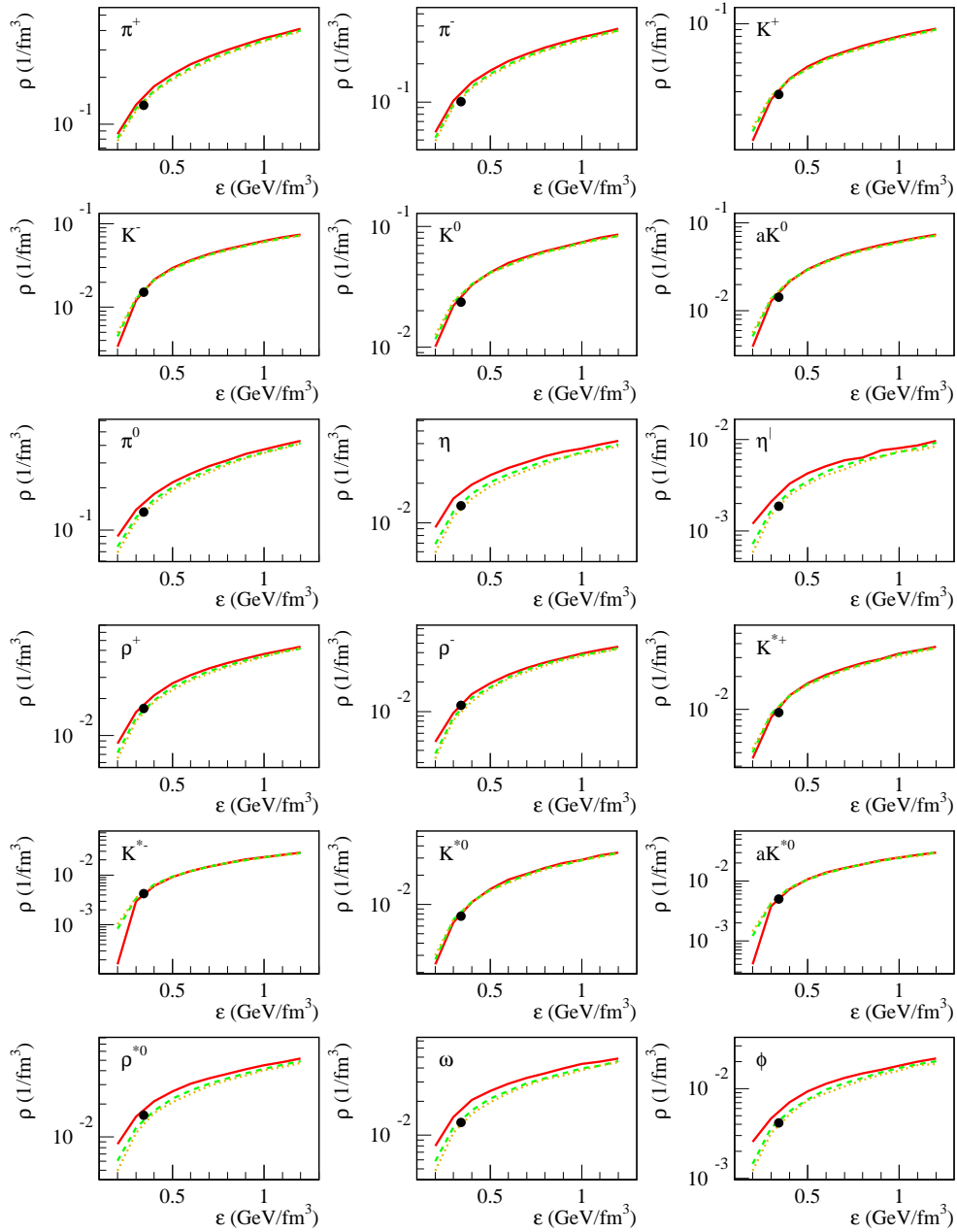}\end{flushleft}

\caption{Density of particles as a function of the energy density of a cluster
for three different volumes: 12.5 $fm^{3}$ (full line), 25 $fm^{3}$
(dashed line) and 50 $fm^{3}$ (dotted line) for an initial baryon
density of 0.08 baryons / $fm^{3}$ and a total charge of 2 using
a microcanonical phase space calculation. The dots present the result
of a canonical calculation provided by Becattini. In both cases the
number of hadrons is limited to 54 and strange particles are not suppressed. }
\end{figure}
\begin{figure}[htb]
\begin{flushleft}\vspace{-2cm}\includegraphics[  scale=0.7]{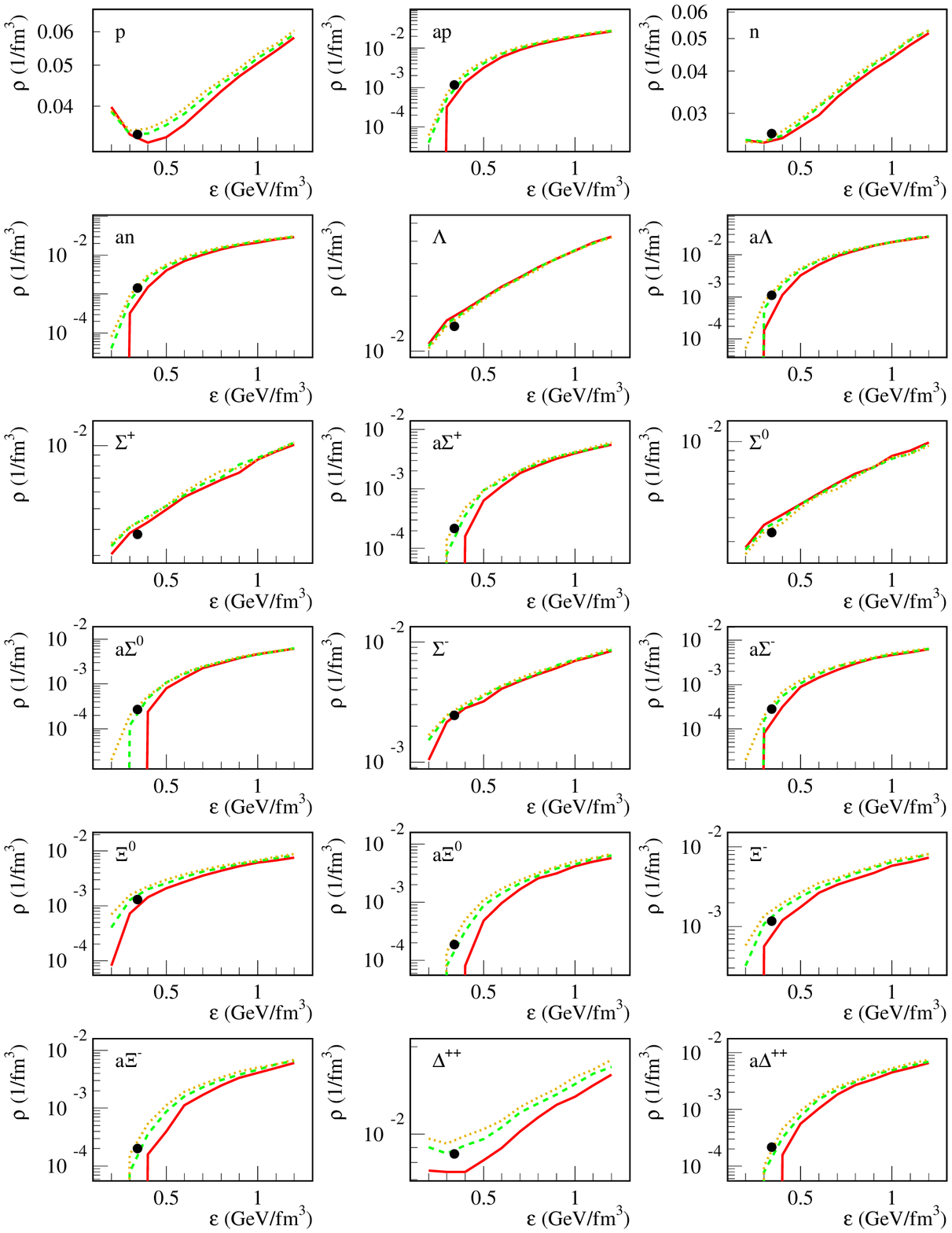}\end{flushleft}

\caption{Same as previous figure, but for additional hadrons}
\end{figure}
\begin{figure}[htb]
\vspace{-2cm}\includegraphics[  scale=0.7]{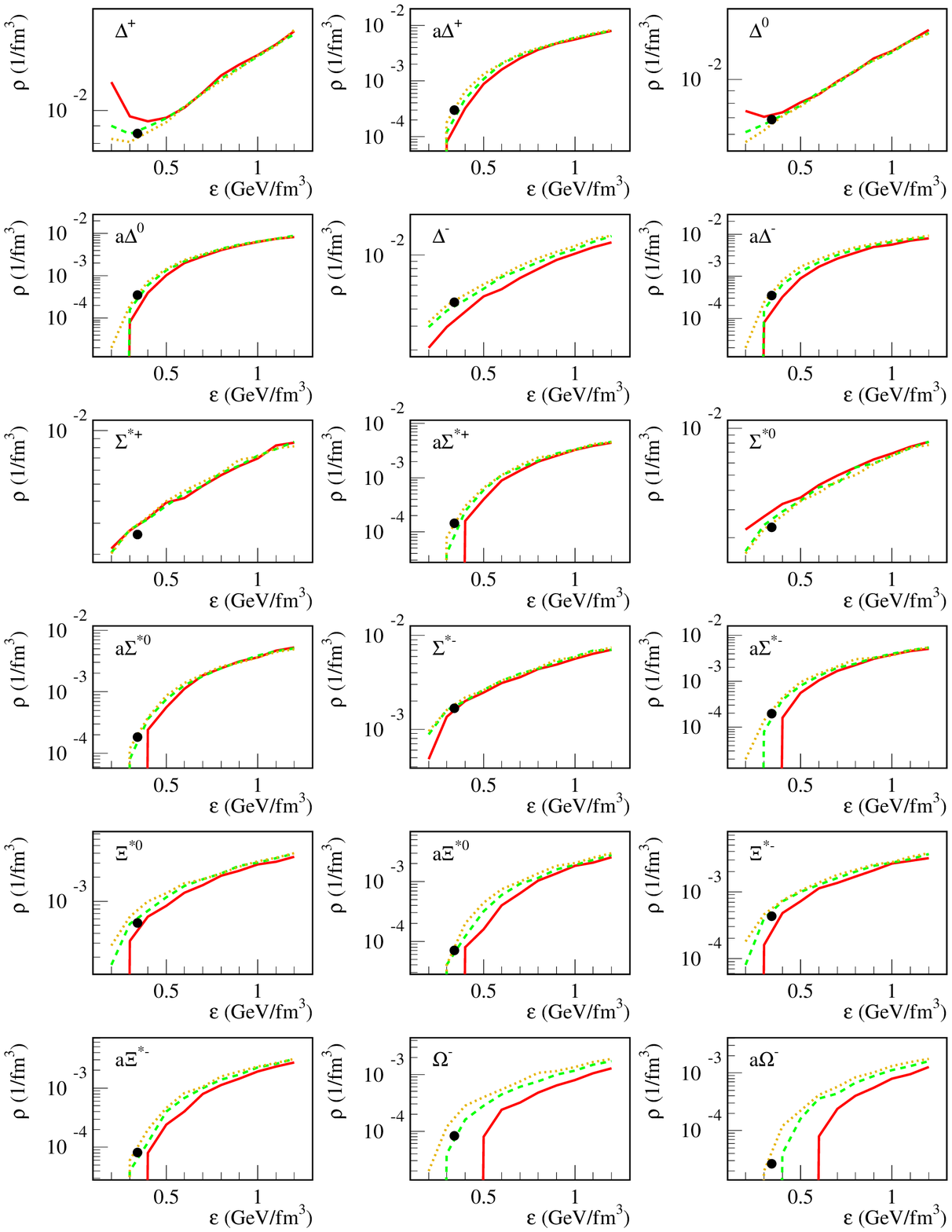}

\caption{Same as previous figure, but for additional hadrons}
\end{figure}
\begin{figure}[htb]
\begin{flushleft}\vspace{-2cm}\includegraphics[  scale=0.7]{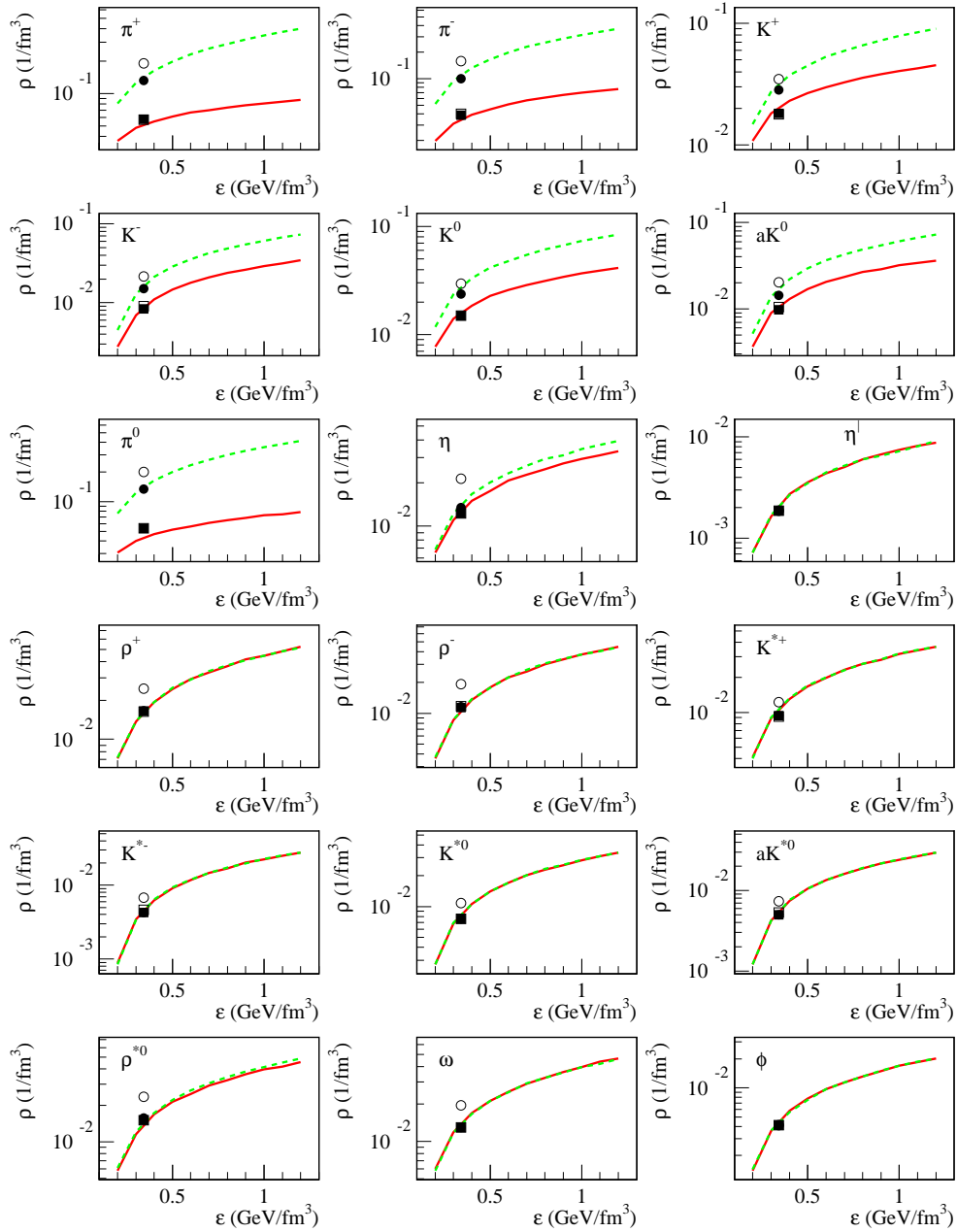}\end{flushleft}

\caption{Density of particles as a function of the energy density of a cluster
with a volume of $50fm^{3}$ in the microcanonical calculation (lines)
in comparison with a canonical calculation at a energy density of
.342 $GeV/fm^{3}$ \label{decnodec} with 54 hadrons (solid points)
and 250 hadrons (open points), before (solid lines and squares) and
after (full lines and circles) strong and electromagneticc decay .}
\end{figure}
\begin{figure}[htb]
\vspace{-2cm}\includegraphics[  scale=0.7]{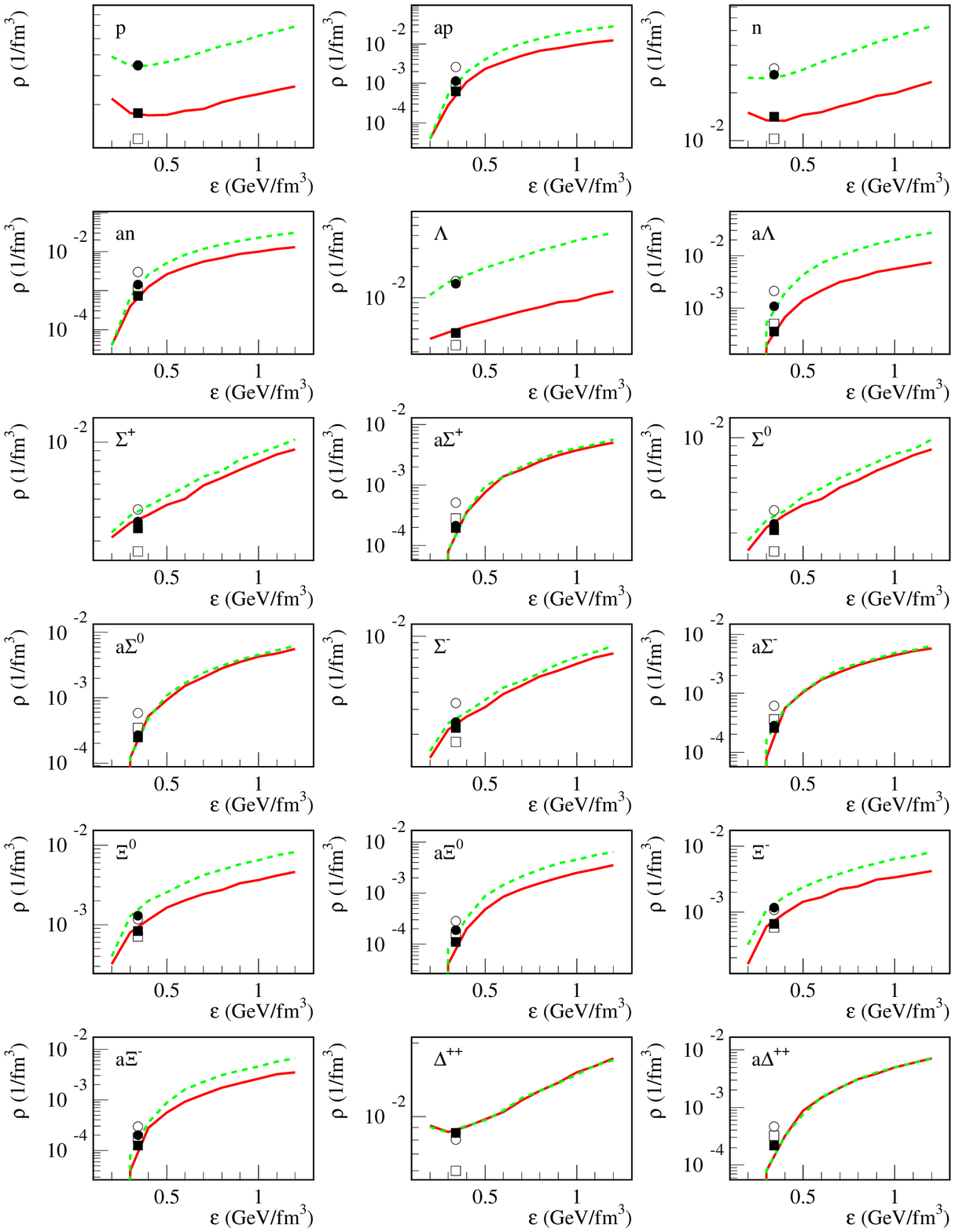}

\caption{Same as previous figure, but for additional hadrons}
\end{figure}
\begin{figure}[htb]
\vspace{-2cm}\includegraphics[  scale=0.7]{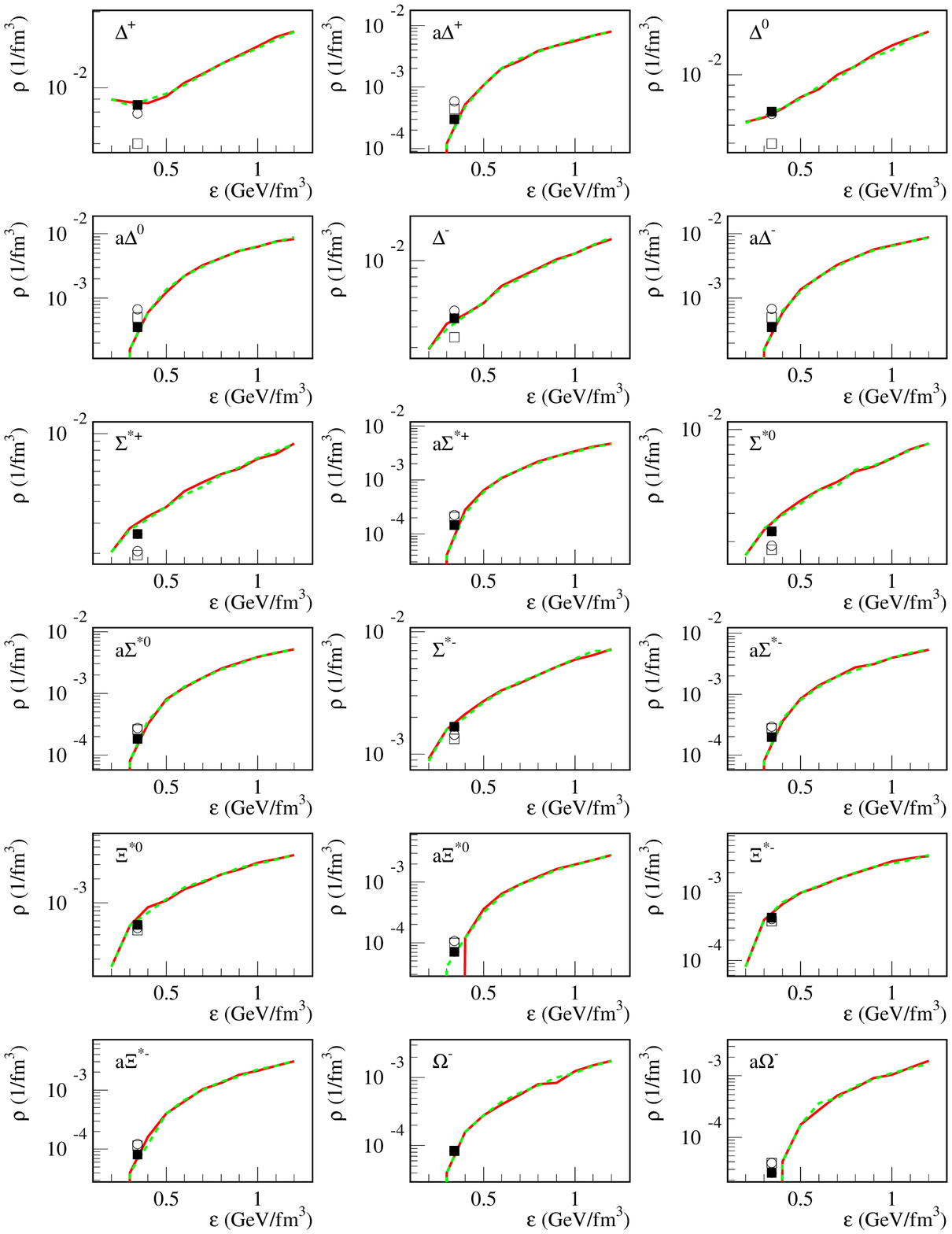}

\caption{Same as previous figure, but for additional hadrons}
\end{figure}

Figs. 1-3 show the average particle density $\rho _{\alpha }$ as
a function of the energy density $\varepsilon $ for three different
volumes V= 12.5 (line), 25 (dashed line) and 50 (dotted) $fm^{3}$
The baryon density is fixed at 0.08 $baryons/fm^{3}$ and the total
charge is 2 for reasons which will be explained later. The particle
densities for V = 12.5 $fm^{3}$ and V = 25 $fm^{3}$ differ at low
energy densities for heavy mesons as well as for many of the baryons.
The deviation increases - as expected - either with increasing particle
mass or with increasing strange quark content, because the strange
quarks have to be compensated by other particles in order to obtain
the quantum numbers of the two incoming protons. 

We see as well that all mesons and almost all of the baryons have
already arrived at the canonical limit for a volume as small as V=
25 $fm^{3}$. The only exceptions are the heavy triple strange barons
$\Omega $ and $\bar{\Omega }$ at low energy densities, where the
total energy of the disintegrating system is (at V= 25$fm^{3}$) only
a couple of times larger than the mass of the $\Omega $ ,$\bar{\Omega }$
pair. 

From this observation we can conclude that the canonical limit is
obtained if the total energy of the system is about 10 times the energy
of the sum of the particles under consideration and of that particles
necessary to compensate the deviation of the quantum numbers of the
considered particle from that of the total cluster. 

With the exception of the proton the density of all particles increases
with increasing energy density. For the proton this is not the case
because at very low energy densities the baryon cannot get rid of
its charge, whereas at higher energies a produced meson can carry
this charge leaving behind a neutral baryon. 

In Figs. 1-3 we have marked as a dot the results of a canonical calculation
with the parameters fitted to describe the particle yields observed
in pp collisions at 27.4 GeV \cite{bec1,becpp}. The parameters used
in this calculations are $V=25.5$ $fm^{3}$ and $T=162$ $MeV$.
In contradistinction to the calculation with these parameters which
is presented in ref. \cite{becpp}, here the strange particles are
not suppressed by a $\gamma _{s}$ factor and only the 54 hadron species
mentioned above are produced. Therefore this canonical calculation
can directly be compared with our microcanonical approach. The average
energy of the clusters obtained in this calculation is 8.74 GeV, resulting
in an average energy density of $\epsilon =0.342$ $GeV/fm^{3}$.
We observe that for all light hadrons the agreement is very nice,
verifying that indeed at a volume of $V=25\, fm^{3}$ the multiplicity
has arrived at its canonical limit. For the heavy particles, such
as $\Omega $ and $\bar{\Omega }$, volumes above $50\, fm^{3}$ are
needed. 

Another interesting question is how the observed particle yield is
influenced by feeding. It is addressed in figs. 4-6 which show for
the 54 hadrons two densities: that of those particles which are produced
directly and that present after electromagnetic and strong decays.
Both microcanonical calculations are compared with the corresponding
canonical results. We see that feeding is unimportant for the vector
mesons, whereas for the pseudoscalar mesons feeding increases the
yield by more than a factor of 2 for the $\pi $'s and by more than
50\% for the K's. With the exceptions of protons, neutrons and $\Lambda $'s
feeding is much less important in the baryonic sector. The strange
(anti)baryons in the electromagnetic decay chain of $\Omega $ and
$\bar{\Omega }$ show some feeding whereas for $\Delta $'s, $\Omega $'s,
$\chi ^{*}$'s and $\Sigma ^{*}$'s feeding is absent. We display
in figs. 3-6 as well the results of the canonical calculation with
the same parameters in which the standard set of hadrons is included.
For the heavy hadrons both canonical results differ little, whereas
the standard set yields more pions and kaons, the decay products of
the hadrons not included in 54 particle set. For the pions both canonical
results differ by less than 50\textbackslash{}\% for all the other
hadrons the differences are much less important.

\subsection{Transverse momentum}

In fig.~7, %
\begin{figure}[htb]
\begin{center}\includegraphics[  scale=0.8]{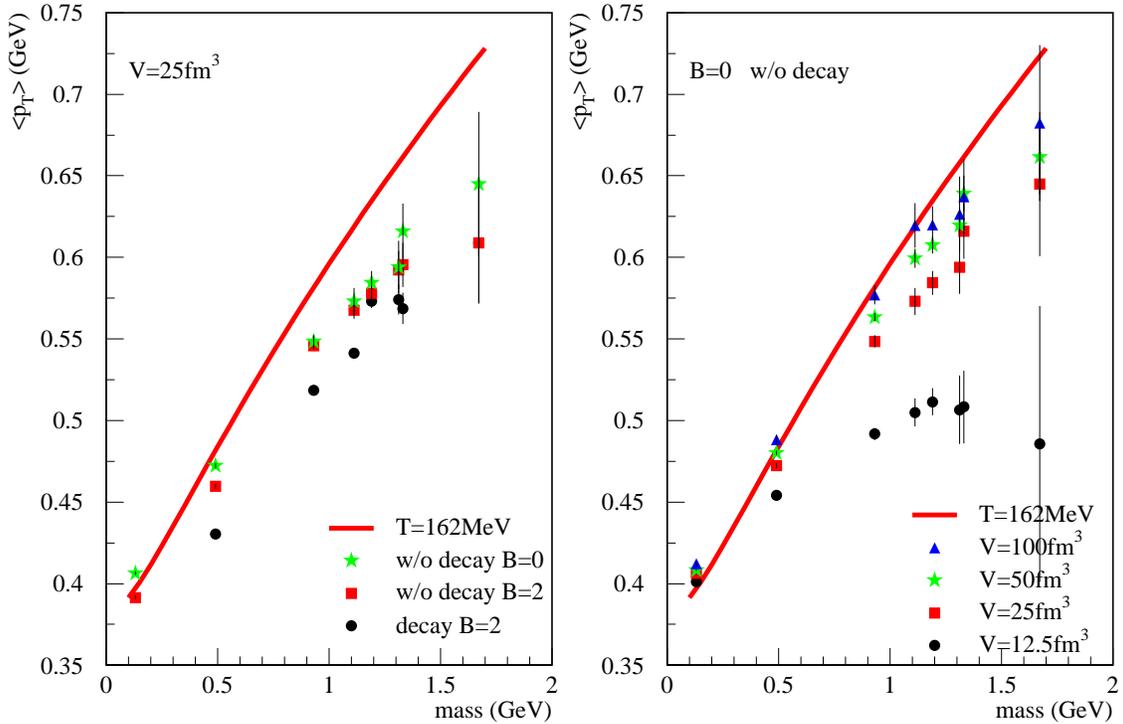}\end{center}

\caption{$<p_{T}(m)>$ for different droplets with an energy density of $0.342\, GeV/fm^{3}$
as compared to the value for the equivalent canonical system (T=162
MeV). On the left hand side we display the results for droplets with
baryon numbers B=0 and B=2, respectively, the latter with and without
strong decay. One the right hand side we present those for a droplet
with different volumes and with a baryon number B=0.}
\end{figure}
we plot the average transverse momentum $<p_{T}>$ as a function of
the hadron mass $m$. We compare our microcanonical results (points)
with an energy density of $\epsilon =0.342\, GeV/fm^{3}$ to the canonical
result, using

\[
<p_{T}(m,T)>=\frac{(\pi mT/2)^{(1/2)}K_{5/2}(m/T)}{K_{2}(m/T)}\]
\cite{hag2,raf}, with a temperature of T = 162 MeV (solid lines).
The $<p_{T}>$values obtained in the canonical calculation of Becattini
agree with that analytical formula.

On the left hand side, we display $<p_{T}>$ of hadrons before decay
as a function of their mass, for a volume of $V=25\, fm^{3}$, and
a total baryon number $B=0$ (stars) and $B=2$ (squares). Whereas
the $<p_{T}>$ of $\pi $'s and K's are close to the value one expects
for an equivalent canonical system, for the heavier particles the
deviation from the canonical value increases with increasing mass.
The reason is easy to understand: If such a heavy particle is produced,
the available energy is not sufficient to fill phase space up to the
high momenta. We see as well that the $<p_{T}>$ of the $\Omega $
is 1.5 time the value of a $\pi $ . Please note that relativity moderates
the increase. In a nonrelativistic calculations we expect that $<p_{T}>$
increases as $\sqrt{m}$. We display on the left hand side as well
the average transverse momenta after the strong electromagnetic interaction
decay for the case of $B=2$ (cycles). The decay modifies considerably
the $<p_{T}>$of the light mesons and baryons.

On the right hand side, we study the critical volume for a droplet
with total baryon number 0 (in order to keep the baryon density constant
when changing the volume). We see that with increasing volume the
$<p_{T}>$ of the heavy baryons approaches only very slowly the canonical
value. However, even at volumes of 100$\, fm^{3}$ the asymptotic
value is not reached. Thus kinematical observables are much more suited
to observe the deviations between canonical and microcanonical systems.

\section{Conclusions}

We have presented the first numerical calculations of the multiplicity
distribution of hadrons assuming that they are distributed according
to microcanonical phase space. In this study we have limited ourselves
to a set of 54 hadrons. The results have been compared with the results
of a canonical system which has the same size and the same average
energy, and agreement has been found. We observe that for the multiplicity
the canonical limit is obtained, if the energy of the system is around
ten times the mass of the observed particle plus that of those particles
which are necessary to compensate its deviation from the quantum numbers
of the cluster. These results raise doubts whether heavy hadrons in
small systems like pp, Kp and $e^{+}e^{-}$ can be described in a
canonical approach but shows as well that for a heavy collision a
canonical treatment is valid. Both, canonical and microcanonical calculations,
show that feeding is very important for the pseudoscalar mesons as
well as for p,n and $\Lambda $ but that is of minor or no importance
for the other particles. We observe, as predicted analytically, an
increase of the average transverse momentum $<p_{T}>$ with the mass
of the hadron. However, we also find that a very big volume is required
to obtain the value of the equivalent canonical system. 

\textbf{Acknowledgment} We would like to thank F. Becattini for many
discussions and for providing us with the calculation with a restricted
set of hadrons and without strangeness suppression.

\end{document}